\documentclass[a4paper]{amsart}
\usepackage{amssymb}
\usepackage{graphics}
\usepackage{vaucanson-g}
\usepackage{slashbox}
\usepackage[all]{xy}

\theoremstyle{plain}
\newtheorem{theorem}{Theorem}
\newtheorem{lemma}[theorem]{Lemma}
\newtheorem{corollary}[theorem]{Corollary}
\newtheorem{proposition}[theorem]{Proposition}
\theoremstyle{definition}
\newtheorem{definition}[theorem]{Definition}

\newtheorem{remark}[theorem]{Remark}

\newtheorem{example}[theorem]{Example}

\DeclareMathOperator{\val}{val}
\DeclareMathOperator{\rep}{rep}
\DeclareMathOperator{\pref}{Pref}

\DeclareMathOperator{\card}{Card}
\DeclareMathOperator{\adh}{Adh}
\DeclareMathOperator{\centre}{Center}

\newcommand{\kom}[1]{}

\title{Representing Real Numbers in a Generalized Numeration System}

\author[E. Charlier]{Emilie Charlier}
\author[M. Le Gonidec]{Marion Le Gonidec}
\author[M. Rigo]{Michel Rigo}

\address[E. Charlier, M. Rigo]{Institute of Mathematics, University of Li\`ege, 
Grande Traverse 12 (B 37), B-4000 Li\`ege, Belgium}
\email{\{echarlier,M.Rigo\}@ulg.ac.be}

\address[M. Le Gonidec]{Laboratoire IMATH, Bat U bureau 308
Avenue de l'université - BP 20132
83957 LA GARDE Cedex
FRANCE Toulon, France}
\email{marion.le-gonidec@univ-tln.fr}

\begin{document}

\begin{abstract}
We show how to represent an interval of real numbers in an abstract 
numeration system built on a language that is not necessarily regular. 
As an application, we consider representations of real numbers using the Dyck language. 
We also show that our framework can be applied to the rational base numeration systems.
\end{abstract}

\maketitle

\section{Introduction}
In \cite{LR2}, P. Lecomte and the third author 
showed how to represent an interval of real numbers 
in an abstract numeration system built 
on a regular language satisfying some suitable conditions. 
In this paper, we provide a wider framework 
and we show that their results can be extended to 
abstract numeration systems built on a language 
that is not necessarily regular. 
Our aim is to provide a unified approach for the 
representation of real numbers in various numeration 
systems encountered in the literature \cite{AFS,DT,LR,Lot}.

This paper is organized as follows.
In the second section, we recall some useful definitions 
and results from automata theory.
In Section~\ref{sec:ANS}, we restate the general framework of \cite{LR2}.
Then in Section~\ref{sec:pref}, we show that the infinite words obtained 
as limits of words of a language are exactly the infinite words 
having all their prefixes in the corresponding prefix closure.
In view of this result, we shall consider 
only abstract numeration systems built 
on a prefix-closed language to represent the reals. 
One can notice that usual numeration systems like integer bas systems, 
$\beta$-numeration or substitutive numeration systems 
are all built on prefix-closed languages \cite{DT,Lot}.
In Section~\ref{sec:reals}, 
we show how to represent an interval $[s_0,1]$ of real numbers 
in a generalized abstract numeration system built 
on a language satisfying some general hypotheses. 
Finally, in Section~\ref{sec:appl}, we give three applications
of our methods, that were not settled yet by the results of \cite{LR2}. 
First, we consider a non-regular language $L$ such that 
its prefix-language $\pref(L)$ is regular.  
In a second part, we consider the representation of real numbers 
in the generalized abstract numeration system 
built on the language of the prefixes of the Dyck words. In this case,
neither the Dyck language $D$ nor its prefix-closure $\pref(D)$ 
are recognized by a finite automaton. 
We compute the complexity functions 
of this language, i.e., 
for each word $w$, the function mapping an integer 
$n$ onto $\card(w^{-1}D\cap \{a,b\}^n)$,
and we show that we can apply our results 
to the corresponding abstract numeration system. 
The third application that we consider is the abstract 
numeration system built on the language $L_{\frac{3}{2}}$ recently introduced in \cite{AFS}.
We show that our method leads, up to some scaling factor, to the same 
representation of the reals as the one given in \cite{AFS}.

\section{Preliminaries}
Let us recall some usual definitions. 
For more details, see for instance \cite{Ei} or \cite{Saka}. 
An {\em alphabet} is a non-empty finite set of symbols, called {\em letters}. 
A {\em word} over an alphabet $\Sigma$ is a finite or infinite 
sequence of letters in $\Sigma.$ 
The {\em empty word} is denoted by $\varepsilon$. 
The set of finite (resp. infinite) words over $\Sigma$ 
is denoted by $\Sigma^*$ (resp. $\Sigma^\omega$). 
The set $\Sigma^*$ is the free monoid generated by $\Sigma$ with respect 
to the concatenation product of words 
and with $\varepsilon$ as neutral element.
A {\em language} (resp. $\omega$-language) over $\Sigma$ 
is a subset of $\Sigma^*$ (resp. $\Sigma^\omega$).
If $w$ is a finite word over $\Sigma$,
the length of $w$, denoted by $|w|$, is the number of its letters 
and if $a\in\Sigma$, then $|w|_a$ is the number of occurrences of $a$ in $w$. 
If $w$ is a finite (resp. infinite) word over $\Sigma$, then
for all $i\in[\![0,|w|-1]\!]$ (resp. $i\in\mathbb{N}$), 
$w[i]$ denotes its $(i+1)$st letter,
for all $0\le i\le j\le |w|-1$ (resp. $0\le i\le j$), 
the {\em factor} $w[i,j]$ of $w$ is the word $w[i]\cdots w[j]$, and 
for all $i\in[\![0,|w|]\!]$ (resp. $i\in\mathbb{N}$), 
$w[0,i-1]$ is the {\em prefix} of length $i$ of $w$, 
where we set $w[0,-1]:=\varepsilon$. 
The set of prefixes of a word $w$ (resp. a language $L$) 
is denoted by $\pref(w)$ (resp. $\pref(L)$).
Notice that indices are counted from $0$.

One can endow $\Sigma^\omega\cup\Sigma^*$ 
with a metric space structure as follows. 
If $x$ and $y$ are two distinct infinite words over $\Sigma$, 
define the distance $d$ over $\Sigma^\omega$ by
$d(x,y):=2^{-\ell}$ 
where $\ell=\inf\{i\in\mathbb{N}\;|\;x[i]\neq y[i]\}$ 
is the length of the maximal common prefix between $x$ and $y$. 
We set $d(x,x)=0$ for all $x\in\Sigma^\omega$.
This distance can be extended to $\Sigma^\omega\cup\Sigma^*$ 
by replacing the finite words $z$ by $z\#^\omega$, 
where $\#$ is a new letter not in $\Sigma$. 
A sequence $(w^{(n)})_{n\ge0}$ of words over $\Sigma$ 
{\em converges} to an infinite word $w$ over $\Sigma$ if 
$d(w^{(n)},w)\to 0$ as $n\to+\infty$.

A deterministic (finite or infinite) automaton over an alphabet $\Sigma$ 
is is a directed graph $\mathcal{A}=(Q,q_0,\Sigma,\delta,F)$, 
where $Q$ is the set of {\em states}, 
$q_0$ is the {\em initial state},  
$F\subseteq Q$ is the set of {\em final states}
and $\delta\colon Q\times \Sigma \to Q$ is the {\em transition function}. 
The transition function can be naturally extended to $Q\times \Sigma^*$ by 
$\delta(q,\varepsilon)=q$ and $\delta(q,aw)=\delta(\delta(q,a),w)$ 
for all $q\in Q$, $a\in \Sigma$ and $w\in\Sigma^*$.  
We often use $q\cdot w$ as shorthand for $\delta(q,w)$.
A state $q\in Q$ is {\em accessible} (resp. {\em coaccessible}) 
if there exists a word $w\in\Sigma^*$ such that $\delta(q_0,w)=q$ 
(resp. $\delta(q,w)\in F$)
and $\mathcal{A}$ is {\em accessible} (resp. {\em coaccessible}) 
if all its state are accessible (resp. coaccessible).
A word $w\in\Sigma^*$ is {\em accepted} 
by $\mathcal{A}$ if $\delta(q_0,w)\in F$. 
The set of accepted words is the {\em language recognized by $\mathcal{A}$}. 
A deterministic automaton is said to be {\em finite} (resp. {\em infinite}) 
if its set of states is finite (resp. infinite).
A language is {\em regular} if it is recognized by some deterministic finite automaton (DFA). 

Among all the deterministic automata recognizing a language, 
one can distinguish the minimal automaton of this language, which is unique up to isomorphism and is defined as follows. 
The {\em minimal automaton} of a language $L$ over an alphabet $\Sigma$ 
is the deterministic automaton $\mathcal{A}_L=(Q_L,q_{0,L},\Sigma,\delta_L,F_L)$ 
where the states are the sets $w^{-1}L=\{x\in\Sigma^*\,\vert\, wx\in L\}$, for any $w\in\Sigma^*$, 
the initial state is $q_{0,L}=\varepsilon^{-1}L=L,$ 
the final states are the sets $w^{-1}L$ with $w\in L$ 
and the transition function $\delta_L$ is defined by 
$\delta_L(w^{-1}L,a)=(wa)^{-1}\,L$ for all $w\in \Sigma^*$ and all $a\in\Sigma$.  
By construction, $\mathcal{A}_L$ is accessible 
and the set of accepted words is exactly $L$. 
It is well known that $\mathcal{A}_L$ is finite if and only if $L$ is regular.
The {\em trim minimal automaton} of a language 
is the minimal automaton of this language 
from which the only possible sink state has been removed, 
i.e. we keep only the coaccessible states. 
In this case, the transition function can possibly be a partial function.

If $L$ is the language recognized by a deterministic automaton $\mathcal{A}=(Q,q_0,\Sigma,\delta,F)$, $L_q:=\{w\in\Sigma^*\,|\, \delta(q,w)\in F\}$ 
is the language of the words accepted from the state $q$ in $\mathcal{A}$ 
and $u_q(n)$ (resp. $v_q(n)$) is the number of words of length $n$ 
(resp. less or equal to $n$) in $L_q$. 
The maps $u_{q}\colon\mathbb{N}\to\mathbb{N}$ are called the 
{\em complexity functions of $\mathcal{A}$}. 
The language $L$ is {\em polynomial} if 
$u_{q_0}(n)$ is $\mathcal{O}(n^k)$ for some non-negative integer $k$ 
and {\em exponential} if 
$u_{q_0}(n)$ is $\Omega(\theta^n)$ for some $\theta>1$, i.e., 
if there exists a constant $c>0$ such that $u_{q_0}(n)\ge c\,\theta^n$ 
for infinitely many non-negative integers $n$.

\section{Generalized Abstract Numeration Systems}\label{sec:ANS}

If $L$ is a language over a totally ordered alphabet $(\Sigma,<)$, 
the {\em genealogical (or radix) ordering $<_{gen}$} over $L$
induced by $<$ is defined as follows.
The words of the language are ordered by increasing length and
for words of the same length, one uses the lexicographical ordering
induced by $<$. 
Recall that for two words $x,y\in \Sigma^*$ of same length,
$x$ is lexicographically less than $y$ if there exist
$w,x',y'\in \Sigma^*$ and $a,b\in \Sigma$ such that $x=wax'$,
$y=wby'$ and $a<b$.
The lexicographical ordering is naturally extended to infinite words.

\begin{definition}
A {\em(generalized) abstract numeration system} is a triple
$S=(L,\Sigma,<~)$ where $L$ is an infinite 
language over a totally ordered alphabet $(\Sigma,<)$. 
Enumerating the words of $L$ using the genealogical order $<_{gen}$ induced by
the ordering $<$ on $\Sigma$ gives a one-to-one correspondence
$\rep_S\colon \mathbb{N}\to L$ mapping the non-negative integer $n$ onto
the $(n+1)$st word in $L$. In particular, $0$ is sent onto the
first word in the genealogically ordered language $L$. 
The reciprocal map is denoted by $\val_S\colon L\to\mathbb{N}$ 
and for all $w\in L$, $\val_S(w)$ is called the 
{\em $S$-numerical value} of $w$.
\end{definition}

Compare with \cite{LR}, we do not ask the language of the numeration to be regular. It is the reason for the introduction of the terminology ``generalized''.

\begin{example}\label{ex:GANS}
Let $\Sigma=\{a,b\}$, 
$L=\{w\in\Sigma^*\colon \left| |w|_a-|w|_b \right|\le 1\}$, and $S=(L,\Sigma,a<b)$. 
The minimal automaton of $L$ is given in Figure~\ref{aut:exGANS}. 
The first words of the $L$ are 
\[\varepsilon,a,b,ab,ba,aab,aba,abb,baa,bab,bba,aabb,abab,abba,baab,\ldots\]

\begin{figure}[htbp]
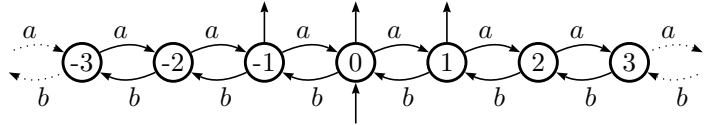

        \centering
\VCDraw{%
        \begin{VCPicture}{(-6,-1)(6,1)}
 \State[0]{(0,0)}{I}
 \State[1]{(2,0)}{R1}
 \State[2]{(4,0)}{R2}
 \State[3]{(6,0)}{R3}
 \State[$-$1]{(-2,0)}{L1}
 \State[$-$2]{(-4,0)}{L2}
 \State[$-$3]{(-6,0)}{L3}
\HideState
\State{(8,0)}{R4}
\State{(-8,0)}{L4}
\Initial[s]{I}
\Final[n]{I}
\Final[n]{R1}
\Final[n]{L1}
\VArcL{arcangle=25}{I}{R1}{a}
\VArcL{arcangle=25}{R1}{R2}{a}
\VArcL{arcangle=25}{R2}{R3}{a}
\VArcL{arcangle=25}{R3}{R2}{b}
\VArcL{arcangle=25}{R2}{R1}{b}
\VArcL{arcangle=25}{R1}{I}{b}
\VArcL{arcangle=25}{I}{L1}{b}
\VArcL{arcangle=25}{L1}{L2}{b}
\VArcL{arcangle=25}{L2}{L3}{b}
\VArcL{arcangle=25}{L3}{L2}{a}
\VArcL{arcangle=25}{L2}{L1}{a}
\VArcL{arcangle=25}{L1}{I}{a}
\ChgEdgeLineStyle{dotted}
\VArcL{arcangle=25}{R3}{R4}{a}
\VArcL{arcangle=25}{R4}{R3}{b}
\VArcL{arcangle=25}{L3}{L4}{b}
\VArcL{arcangle=25}{L4}{L3}{a}
\end{VCPicture}
}
        \caption{The minimal automaton of $L$.}
        \label{aut:exGANS}
\end{figure}
\end{example}

The following proposition is a result from \cite{LR2} 
extended to any language.
This shows how to compute the numerical 
value of a word in the numeration language. 

\begin{proposition}\label{lem:val}
Let $S=(L,\Sigma,<)$ be a (generalized) abstract numeration system 
and let $\mathcal{A}=(Q,q_0,\Sigma,\delta,F)$ 
be a deterministic automaton recognizing $L$. 
If $w\in L$, then we have
\[\val_S(w)=v_{q_0}(|w|-1)
+\sum_{i=0}^{|w|-1}\sum_{a<w[i]}u_{q_0\cdot w[0,i-1]a}(|w|-i-1).\]
\end{proposition}

\section{Languages $L$ with Uncountable $\adh(L)$} \label{sec:pref}

The notion of adherence has been introduced in \cite{N} 
and has been extensively studied in \cite{BN}. 

\begin{definition}
Let $L$ be a language over an alphabet $\Sigma$.
The {\em adherence of $L$}, denoted by $\adh(L)$,
is the set of infinite words over $\Sigma$ 
whose prefixes are prefixes of words in $L$:
\[\adh(L)=\{w\in\Sigma^\omega\;|\;\pref(w)\subseteq\pref(L)\}.\] 
\end{definition}

Notice that $\adh(L)$ is empty if and only if $L$ is finite.

For the usual topology on $\Sigma^*\cup\Sigma^\omega$, 
the closure $\bar{L}$ of a language $L$ over $\Sigma$ 
satisfies the equality: $\bar{L}=L\cap\adh(L)$.

The following lemma gives a characterization of the adherence of a language \cite{BN}. 
We give a proof for the sake of completeness.

\begin{lemma}\label{lem:adh}
Let $L$ be a language over an alphabet $\Sigma$.
The adherence of $L$ 
is the set of infinite words over $\Sigma$ that are limits of words in $L$: 
\[\adh(L)=\{w\in\Sigma^\omega\;|\;
\exists (w^{(n)})_{n\ge0}\in L^\mathbb{N},\, w^{(n)}\to w\}.\]
\end{lemma}

\begin{proof}
Take an infinite word $w$ in $\adh(L)$.
Then for all $n\ge0$, we have $w[0,n-1]\in\pref(L)$. 
Thus for all $n\ge0$, there exists a finite word $z^{(n)}\in\Sigma^*$ 
such that $w^{(n)}:=w[0,n-1]z^{(n)}$ belongs to $L$. 
Obviously $w^{(n)}\to w$ and $w$ belongs to the r.h.s. set in the statement.
Conversely, take an infinite word $w$ which is the limit of a sequence 
$(w^{(n)})_{n\ge0}$ of words in $L$. 
Then for all $\ell\ge0$, there exists $n\ge0$ such that 
we have $w[0,\ell-1]\in\pref(w^{(n)})\subseteq\pref(L)$. 
This shows that $w$ belongs to~$\adh(L)$.
\end{proof}

The notion of center of a language can be found in \cite{BN}. 

\begin{definition}
Let $L$ be a language over an alphabet $\Sigma$.
The {\em center of $L$}, denoted by $\centre(L)$, 
is the prefix-closure of the adherence of $L$:
\[\centre(L)=\pref(\adh(L)).\]
\end{definition}

The next lemma gives a characterization of the center of a language \cite{BN}. 
Again we give a proof for the sake of completeness.

\begin{lemma}\label{lem:centre}
Let $L$ be a language over an alphabet $\Sigma$.
The center of $L$ is the set of words which are prefixes of an infinite number 
of words in $L$:
\[\centre(L)=\{w\in\pref(L)\;|\;w^{-1}L \text{ is infinite}\}.\]
\end{lemma}

\begin{proof}
Take a word $w$ in $\centre(L)$.
By defnition, 
there exists a infinite word $z$ over $\Sigma$ such that
$wz$ belongs to $\adh(L)$. 
Then for all $n\ge0$, $wz[0,n-1]$ belongs to $\pref(L)$. 
Thus for all $n\ge0$, there exists a finite word $y^{(n)}\in\Sigma^*$ 
such that $w^{(n)}:=wz[0,n-1]y^{(n)}$ belongs to $L$, 
and there are infinitely many such words $w^{(n)}$. 
Conversely, let $w$ be a prefix of infinitely many words in $L$. 
There exists a letter $a\in\Sigma$ such that $wa$ is a prefix of infinitely many words in $L$. 
Iterating this argument, there exists a sequence $(a_n)_{n\ge0}$ of letters in $\Sigma$ 
such that $wa_0\cdots a_n$ belongs to $\pref(L)$ for all $n\ge0$. 
This implies that $wa_0a_1 \cdots$ belongs to $\adh(L)$. 
Hence $w$ belongs to $\centre(L)$.
\end{proof}

\begin{definition}
If $L$ is a language over an alphabet $\Sigma$, 
\[L_\infty=\{w\in\Sigma^\omega\;|\;
\exists^\infty n\in\mathbb{N},\, w[0,n-1]\in L\}\] 
denotes the set of infinite words over $\Sigma$ 
having infinitely many prefixes in $L$.
\end{definition}

Again, observe that $L_\infty$ is empty if and only if $L$ is finite.

The following lemma is obvious.

\begin{lemma}\label{lem:pref1}
For any language $L$, we have $L_\infty\subseteq\adh(L)$.
Moreover, if $L$ is a prefix-closed language, then $L_\infty =\adh(L)$.
\end{lemma}

Let us recall two results from \cite{LR2}. 

\begin{proposition}\label{prop:LR2} 
Let $L$ be a regular language.
The set $\adh(L)$ is uncountably infinite 
if and only if, in any deterministic finite automaton accepting $L$, 
there exist at least
two distinct cycles $(p_1,\ldots,p_r,p_1)$ 
and $(q_1,\ldots,q_s,q_1)$ where $r,s\ge 2$, 
starting from the same accessible and coaccessible state $p_1=q_1$.
\end{proposition}

\begin{proposition}\label{prop:ldinfty}
Let $L$ be a regular language.
The set $L_\infty$ is uncountably infinite 
if and only if, in any deterministic finite automaton accepting $L$, 
there exist at least
two distinct cycles $(p_1,\ldots,p_r,p_1)$ and $(q_1,\ldots,q_s,q_1)$ where $r,s\ge 2$, 
starting from the same accessible state $p_1=q_1$ 
and such that each of them contains at least a final state.
\end{proposition}

It is well known \cite{Sz} that the set of regular 
languages splits into two parts: 
the set of exponential languages and the set of polynomial languages. 
The polynomial regular languages over an alphabet $\Sigma$ are exactly those 
that are finite union of languages of the form 
\begin{equation}\label{eq:polynomial}
x_1y_1^*x_2y_2^*\cdots x_ky_k^*x_{k+1}
\end{equation}
where $k\ge 0$ and the $x_i$'s and the $y_i$'s are finite words over $\Sigma$.
Consequently, in view of Proposition~\ref{prop:LR2}, 
the following result is obvious.

\begin{corollary}\label{cor:regexp}
If $L$ is a regular language, then
the following assertions are equivalent:
\begin{itemize}
\item $\adh(L)$ is an uncountable set;
\item $L$ is exponential;
\item $\pref(L)$ is exponential. 
\end{itemize}
\end{corollary}

If the considered language is not regular, 
then only the sufficient conditions of 
Proposition~\ref{prop:LR2} and Proposition~\ref{prop:ldinfty} hold true.
They can be reexpressed as follows.

\begin{proposition}\label{prop:linfty} 
If, in any deterministic automaton accepting a language $L$, there exist at least
two distinct cycles $(p_1,\ldots,p_r,p_1)$ and $(q_1,\ldots,q_s,q_1)$ where $r,s\ge 2$, 
starting from the same accessible and coaccessible state $p_1=q_1$, 
then the set $\adh(L)$ is uncountably infinite and $L$ is exponential.
\end{proposition}

\begin{proposition}\label{prop:linfty2} 
If, in any deterministic automaton accepting a language $L$, there exist at least
two distinct cycles $(p_1,\ldots,p_r,p_1)$ and $(q_1,\ldots,q_s,q_1)$ where $r,s\ge 2$, 
starting from the same accessible state $p_1=q_1$ 
and such that each of them contains at least a final state, 
then the set $L_\infty$ is uncountably infinite and $L$ is exponential.
\end{proposition}

There exist non-regular exponential languages 
with an uncountable associated set $L_\infty$, 
and thus also with an uncountable set $\adh(L)$,
that are recognized by a deterministic automaton without 
distinct cycles satisfying condition of Proposition~\ref{prop:linfty}.
For instance, see Example~\ref{ex:3/2} of Section~\ref{sec:appl} 
about the $\frac{3}{2}$-number system.
Notice that the corresponding trim minimal automaton depicted in Figure~\ref{aut:3/2} 
has an infinite number of final states. 
Note that, by considering automata having a finite set of final states, 
we get back the necessary condition of Proposition~\ref{prop:ldinfty}.

\begin{proposition}
Let $L$ be a language recognized by a 
deterministic automaton $\mathcal{A}$ having a finite set of final states. 
The set $L_\infty$ is uncountably infinite 
if and only if there exist in~$\mathcal{A}$ at least
two distinct cycles $(p_1,\ldots,p_r,p_1)$ 
and $(q_1,\ldots,q_s,q_1)$ where $r,s\ge 2$, 
starting from the same accessible state $p_1=q_1$ 
and such that each of them contains at least a final state.
\end{proposition}

\begin{proof}
In view of Proposition~\ref{prop:linfty}, 
we only have to show that the condition is necessary.
Since there is only a finite number of final states, 
if $w\in L_\infty$, then there exist a final state~$f$ 
and infinitely many $n$ such that $q_0\cdot w[0,n-1]=f$. 
If $\mathcal{A}$ does not contain such distinct cycles,  
then this implies that any word in $L_\infty$ is of the form 
$xy^\omega$, where $x,y$ are finite words.
Since there is a countable number of such words, we would get that
$L_\infty$ is a countable set. The conclusion follows.
\end{proof}

\begin{corollary}
Let $L$ be a language recognized by a deterministic automaton $\mathcal{A}$ 
having a finite set of final states. 
If $L_\infty$ is an uncountable set, then $L$
is exponential. 
\end{corollary}

\begin{remark}
Any deterministic automaton recognizing a non-regular prefix-closed language 
has an infinite number of final states. 
Indeed, in such an automaton, 
all coaccessible states are final. 
\end{remark}

There exist exponential (and prefix-closed) 
languages $L$ with a countable, and even finite, set $\adh(L)$. 
We give an example of such a language.

\begin{example}
Let $L=\{w\in\{a,b\}^*\;|\;\exists u\in\{a,b\}^*\,\colon\, 
w=a^{\lfloor \frac{|w|}{2}\rfloor}u\}$. 
We have 
\[u_L(n)=  \left\{
\begin{array}{l}
2^{\frac{n}{2}}\;\text{ if } n\equiv 0\mod 2 ,\\
2^{\frac{n+1}{2}}\;\text{ if } n\equiv 1\mod 2
\end{array}
\right. \]
and $\adh(L)=L_\infty=\{a^\omega\}$.
The minimal automaton of $L$ is depicted in Figure~\ref{aut:den}. 
\begin{figure}[htbp]
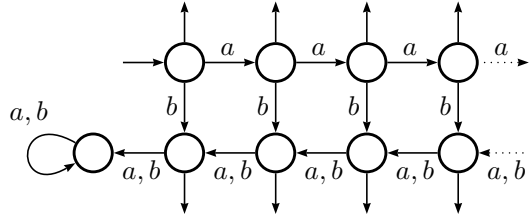

        \centering
\VCDraw{%
        \begin{VCPicture}{(-2,1)(6,5)}
 \State{(0,4)}{I0}
 \State{(0,2)}{Q0}
 \State{(-2,2)}{P}
 \State{(2,4)}{I1}
 \State{(2,2)}{Q1}
 \State{(4,4)}{I2}
 \State{(4,2)}{Q2}
 \State{(6,4)}{I3}
 \State{(6,2)}{Q3}
\HideState
\State{(8,2)}{Q4}
\State{(8,4)}{I4}
\Initial[w]{I0}
\Final[n]{I0}
\Final[n]{I1}
\Final[n]{I2}
\Final[n]{I3}
\Final[s]{Q0}
\Final[s]{Q1}
\Final[s]{Q2}
\Final[s]{Q3}
\EdgeL{I0}{I1}{a}
\EdgeL{I1}{I2}{a}
\EdgeL{I2}{I3}{a}
\EdgeR{I0}{Q0}{b}
\EdgeR{I1}{Q1}{b}
\EdgeR{I2}{Q2}{b}
\EdgeR{I3}{Q3}{b}
\EdgeL{Q3}{Q2}{a,b}
\EdgeL{Q2}{Q1}{a,b}
\EdgeL{Q1}{Q0}{a,b}
\EdgeL{Q0}{P}{a,b}
\LoopW{P}{a,b}
\ChgEdgeLineStyle{dotted}
\EdgeL{I3}{I4}{a}
\EdgeL{Q4}{Q3}{a,b}
\end{VCPicture}
}
        \caption{The minimal automaton of $L$.}
\label{aut:den}
\end{figure}
\end{example}

\section{Representation of Real Numbers}\label{sec:reals}

In the framework of \cite{LR2}, a real number is represented in 
an abstract numeration system built on a regular language $L$ 
as a limit of a sequence of words of~$L$. 
Observe that in this context, thanks to Lemma~\ref{lem:adh}, 
the set of possible representations of the considered reals is $\adh(L)$. 
Therefore, one could consider abstract numeration systems 
built on the prefix-language instead of the one built on the language itself, 
see Remark~\ref{rem:justif} and Remark~\ref{rem:justif2}.  
This point of view is relevant if we compare this with the framework
of the classical integer base $b\ge2$ numeration systems. 
Indeed, in these systems, the numeration language is 
\[\mathcal{L}_b:=\{1,2,\ldots,b-1\}\{0,1,\ldots,b-1\}^*,\]
which is of course a prefix-closed language. 
Notice that this is also the case for non-standard numeration systems like $\beta$-numeration systems and substitutive numeration systems.
Adopting this new framework, we consider only abstract numeration systems 
built on prefix-closed languages. 
Therefore, to represent real numbers, we do not distinguish anymore 
abstract numeration systems built on two distinct languages $L$ and $M$ 
such that $\pref(L)=\pref(M)$. 

Let $S=(L,\Sigma,<)$ be a generalized abstract numeration system 
built on a prefix-closed language $L$.
Let $\mathcal{A}=(Q,q_0,\Sigma,\delta,F)$ 
be an accessible deterministic automaton recognizing $L$.
We make the following assumptions:
\medskip

\noindent
{\bf Hypotheses.}
\begin{enumerate}
\item[(H1)] The set $\adh(L)$ is uncountable;
\item[(H2)] $\forall w\in\Sigma^*,\;\exists r_{w}\ge0\colon$ 
                $\lim_{n\to+\infty} \frac{u_{q_0\cdot w}(n-|w|)}{v_{q_0}(n)}=r_w$;      
\item[(H3)] $\forall w\in \adh(L)$, $\lim_{\ell\to+\infty}r_{w[0,\ell-1]}= 0$.
\end{enumerate}

Observe that for all $w\not\in \centre(L)$, we have $r_w=0$. 

Recall that, since $L$ is a prefix-closed language, we have $\adh(L)=L_\infty$, 
see Lemma~\ref{lem:pref1}.

\medskip

\noindent
{\bf Notation.}
We set $r_0:=r_\varepsilon$ and 
\[s_0:=1-r_0=\lim_{n\to+\infty} \frac{v_{q_0}(n-1)}{v_{q_0}(n)}.\]
\smallskip

\begin{remark}
In \cite{LR2} are considered regular languages $L$ 
with uncountably infinite $\adh(L)$ such that, 
for each state $q$ of a DFA recognizing~$L$, either $L_q$ is finite, 
or $u_q(n)\sim P_q(n)\theta_q^n$ where $P_q\in\mathbb{R}[X]$ and $\theta_q\ge1$.
One can notice that such languages satisfy the hypotheses (H1), (H2) and (H3) above.
Indeed, for all states $q$ and all $\ell\ge0$, it can be shown that
\[\lim_{n\to+\infty}\frac{u_q(n-\ell)}{v_{q_0}(n)}=
\frac{a_q\,(\theta_{q_0}-1)}{\theta_{q_0}^{\ell+1}}\]
where $\theta_{q_0}>1$ and $a_q:=\lim_{n\to+\infty}\frac{u_q(n)}{u_{q_0}(n)}$. 
Since $Q$ is finite, this is sufficient to verify our assumptions. 
Notice also that  
for the integer base $b$ numeration system, 
the three hypotheses are trivially satisfied.
\end{remark}
 
We shall represent real numbers 
by infinite words $w$ of $\adh(L)$
by considering the corresponding limit  
\begin{equation}\label{eq:lim}
\lim_{n\to+\infty}\frac{\val_S(w[0,n-1])}{v_{q_0}(n)}.
\end{equation} 

Our aim is to show that for all $w\in \adh(L)$, 
the limit \eqref{eq:lim} exists, see Proposition~\ref{prop:limval}.

\begin{remark}\label{rem:justif}
If the considered abstract numeration system is 
built on a language that is not prefix-closed, 
we cannot guarantee that the limit~\eqref{eq:lim} exists.
Consider for instance the abstract numeration system 
built on the language $L$ of Example~\ref{ex:GANS}, which is not prefix-closed. 
The sequences $((ab)^n)_{n\ge0}$ and $((ab)^na)_{n\ge0}$ 
of words in $L$ converge to the same infinite word $(ab)^\omega$, 
but the corresponding numerical sequences do not converge 
to the same real number. 
More precisely, using notation of Example~\ref{ex:GANS}, we have
\begin{equation}\label{eq:nolim}
\lim_{n\to+\infty}
\frac{\val_S((ab)^n)}{v_0(2n)}=\frac{3}{4}
\;\text{ and }\; 
\lim_{n\to+\infty}
\frac{\val_S((ab)^na)}{v_0(2n+1)}=\frac{3}{5}, 
\end{equation}
so that the limit 
\[\lim_{n\to+\infty}\frac{\val_S((ab)^\omega[0,n-1])}{v_0(n)}\]
does not exist. 
This essentially comes from the staircase behaviour of $(u_0(n))_{n\ge0}$. 
We have that for all $n\ge0$,
\[u_{0}(n)=
 \left\{
\begin{array}{ll}
\binom{n}{\frac{n}{2}} & \text{ if } n\equiv0\mod 2,\\
2\binom{n}{\frac{n-1}{2}}& \text{ if }n\equiv1\mod 2.
\end{array}
\right.
\]
This implies in particular that $\lim_{n\to+\infty} \frac{v_0(n-1)}{v_0(n)}$ does not exist. 
Indeed, using Stirling formula and \cite[Ch. V.4, Prop. 2]{bourbaki}, we have
\begin{equation}\label{eq:vn}
v_0(2n)\sim\frac{8}{3\sqrt{\pi}}n^{-\frac{1}{2}}4^n\; \text{ and }\;
v_0(2n-1)\sim\frac{5}{3\sqrt{\pi}}n^{-\frac{1}{2}}4^n\; (n\to +\infty).
\end{equation}
Hence, 
\[\lim_{n\to+\infty}\frac{v_0(2n-1)}{v_0(2n)}=\frac{5}{8}\; \text{ and }\; \lim_{n\to+\infty}\frac{v_0(2n)}{v_0(2n+1)}=\frac{2}{5}.\]
By Proposition~\ref{lem:val}, we obtain that for all $n\ge1$,
\begin{gather*}
 \frac{\val_S((ab)^n)}{v_0(2n)}=\frac{v_0(2n-1)}{v_0(2n)}+\frac{\sum_{i=0}^{n-1} u_2(2i)}{v_0(2n)},\\
\frac{\val_S((ab)^na)}{v_0(2n+1)}=\frac{v_0(2n)}{v_0(2n+1)}+\frac{\sum_{i=0}^{n-1} u_2(2i+1)}{v_0(2n+1)}.
\end{gather*}
Using again Stirling formula, we get 
\begin{gather*}
u_2(2i)=\binom{2i}{i-1}\sim \frac{1}{\sqrt{\pi}}i^{-\frac{1}{2}}4^{i}\;(i\to +\infty),\\
u_2(2i+1)=\binom{2i+1}{i}+\binom{2i+1}{i-1} \sim \frac{4}{\sqrt{\pi}}i^{-\frac{1}{2}}4^{i}\; (i\to +\infty).
\end{gather*}
Therefore, by \cite[Ch. V.4, Prop. 2]{bourbaki} and in view \eqref{eq:vn}, it follows that
\[\lim_{n\to+\infty}\frac{\sum_{i=0}^{n-1}u_2(2i)}{v_0(2n)}=\frac{1}{8}
\;\text{ and }\;
\lim_{n\to+\infty}\frac{\sum_{i=0}^{n-1}u_2(2i+1)}{v_0(2n+1)}=\frac{1}{5}.\]
and we obtain the limits of \eqref{eq:nolim}. 
\end{remark}

\begin{remark}\label{rem:justif2}
Considering prefix-closed languages not only avoids 
numerical convergence problems as in Remark~\ref{rem:justif}  but also permits to get 
rid of problems arising from languages $L$ such that there is 
infinitely many $n$ for which $L\cap \Sigma^n=\emptyset$ as discussed in \cite[Remark 4]{LR2}.
\end{remark}

\begin{definition}
If $w\in \adh(L)$ is such that 
$\lim_{n\to+\infty}\frac{\val_S(w[0,n-1])}{v_{q_0}(n)}=x$, 
we say that $w$ is an {\em $S$-representation} of $x$. 
\end{definition}

\begin{example}\label{ex:num}
Consider the abstract numeration system built on the Dyck language that will be described in Example \ref{ex:dyck}. Table \ref{tab:num} gives some numerical approximations. We will see further that $\lim_{n\to+\infty}\frac{\val_S((aab)^\omega[0,n-1])}{v_{q_0}(n)}=\frac{39}{49}=0.79592\cdots$.

\begin{table}[htp]
$\begin{array}{l|c|c|c}
w & \val_S(w) & v_{q_0}(|w|) & \frac{\val_S(w)}{v_{q_0}(|w|)}\\
\hline
a & 1 & 2 & 0.50000\\
aa & 2 & 4 & 0.50000\\
aab & 5 & 7 & 0.71429\\ 
aaba & 9 & 13 & 0.69231\\
aabaa & 17 & 23 & 0.73913\\
aabaab & 32 & 43 & 0.74419\\
aabaaba & 60 & 78 & 0.76923\\
aabaabaa & 112 & 148 & 0.75676\\
aabaabaab & 213 & 274 & 0.77737\\
aabaabaaba & 404 & 526 & 0.76806\\
aabaabaabaa & 771 & 988 & 0.78036\\
aabaabaabaab & 1479 & 1912 & 0.77354\\
aabaabaabaaba & 2841 & 3628 & 0.78308\\
aabaabaabaabaa & 5486 & 7060 & 0.77705\\
aabaabaabaabaab & 10591 & 13495 & 0.78481\\
\vdots & \vdots &\vdots &\vdots\\
\end{array}$ 
\caption{Some numerical approximations.}
\label{tab:num}
\end{table}

\begin{figure}[htp]
\begin{center}
\includegraphics{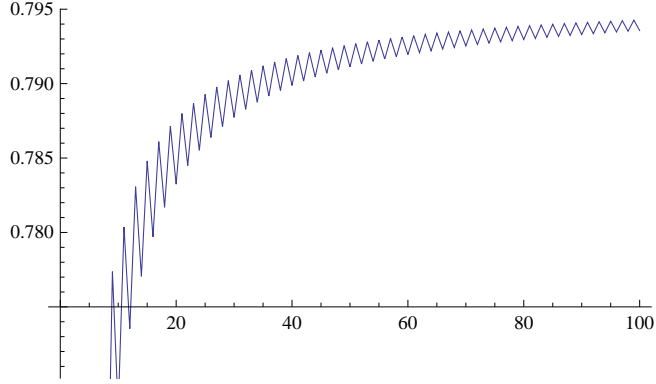}
\caption{The first 100 values of $\frac{\val_S((aab)^\omega[0,n-1])}{v_{q_0}(n)}$.}
\end{center}
\end{figure}
\end{example}

Notice that for all $w\in \adh(L)$, we have
$\val_S(w[0,n-1])\in[v_{q_0}(n-1),v_{q_0}(n)-1]$ for all $n\ge1$. 
Therefore, the represented real numbers $x$ must belong to the interval $[s_0,1]$.
\medskip

Like in \cite{LR2}, we divide $[s_0,1]$ 
into subintervals $I_y$, 
for all prefixes $y$ of infinitely many words in $L$.
For each $\ell\ge0$, $\centre(L)\cap\Sigma^\ell$ is the set
of words of length $\ell$ which are prefixes of infinitely many words of $L$. 
For each $y\in \centre(L)\cap\Sigma^\ell$ and $n\ge\ell\ge0$, 
define
\[\alpha_{y,n}:=\frac{v_{q_0}(n-1)}{v_{q_0}(n)}
+\sum_{\substack{x<y\\x\in \centre(L)\cap\Sigma^\ell}}
\frac{u_{q_0\cdot x}(n-\ell)}{v_{q_0}(n)}\]
and
\[I_{y,n}:=\left[\alpha_{y,n},\alpha_{y,n}
+\frac{u_{q_0\cdot y}(n-\ell)}{v_{q_0}(n)}\right].\]
Then, in view of Hypothesis (H2), 
for all $y\in \centre(L)\cap\Sigma^\ell$, we can define the limit interval
\[I_y:=\lim_{n\to+\infty}I_{y,n}=[\alpha_y,\alpha_y+r_y],\] 
where 
\[\alpha_y:=\lim_{n\to+\infty}\alpha_{y,n}=
s_0+\sum_{\substack{x<y\\x\in \centre(L)\cap\Sigma^\ell}} r_x.\]
Moreover, we set $I_y:=\emptyset$ 
for all $y\in L\setminus\centre(L)$. 
From \cite{LR2}, we know that for all $\ell\ge0$, we have 
\[[s_0,1]=\bigcup_{y\in \centre(L)\cap\Sigma^\ell}I_y\] 
and for all $y,z\in\Sigma^*$, 
\begin{equation}\label{eq:I}
I_{yz}\subseteq I_y. 
\end{equation}
More precisely, if $a_1,\ldots,a_k$ are the letters of $\Sigma$ 
and if $a_1<\cdots<a_k$, then for all $y\in \centre(L)$ 
and all $j\in[\![1,k]\!]$ such that $ya_j\in \centre(L)$, one has 
\begin{equation}\label{I_ya}
I_{ya_j}=
\left[\alpha_y+\sum_{i=1}^{j-1} r_{ya_i},
\alpha_y+\sum_{i=1}^{j} r_{ya_i}\right].
\end{equation}

\begin{remark}
Let $y,z$ be words in $\Sigma^*$ such that $yz\in L$.
If $y$ is prefix of infinitely many words in $L$ 
and if $|z|$ is large enough so that every word 
of length $|yz|$ has a prefix in $\centre(L)\cap\Sigma^{|y|}$, 
then we have
\begin{equation}\label{eq:valpref}
\val_S(yz)=v_{q_0}(|yz|-1)+\sum_{\substack{x<y\\x\in \centre(L)\cap\Sigma^{|y|}}} u_{q_0\cdot x}(|z|)+\sum_{i=|y|}^{|yz|-1} \sum_{\substack{x<yz[0,i]\\|x|=i+1}} u_{q_0\cdot x}(|yz|-i-1).
\end{equation}
\end{remark}

\begin{lemma}\label{lem:a_w}
Let $w\in \adh(L)$. For all $\ell\ge0$, 
$w[0,\ell-1]$ belongs to $\centre(L)\cap\Sigma^\ell$ and the limit 
\[\lim_{\ell\to+\infty}\alpha_{w[0,\ell-1]}\]
exists.
\end{lemma}

\begin{proof}
The first part is obvious 
since $w[0,\ell-1]$ is a prefix of $w[0,n-1]$ for any $n\ge\ell$, 
see Lemma \ref{lem:centre}. 
For the second part, on the one hand, 
observe that \eqref{eq:I} implies that for all $\ell\ge1$,
$\alpha_{w[0,\ell-1]}\le\alpha_{w[0,\ell]}$. 
On the other  hand, we have also that for all $\ell\ge1$,
$\alpha_{w[0,\ell-1]}\le 1$.
Hence, $(\alpha_{w[0,\ell-1]})_{\ell\ge1}$
is a bounded and non-decreasing sequence, so it must converge. 
\end{proof}

\noindent
{\bf Notation.}
For all $w\in \adh(L)$, 
$\alpha_w:=\lim_{\ell\to+\infty}\alpha_{w[0,\ell-1]}$. 
\medskip

Note that we have $\alpha_w\ge\alpha_{w[0,\ell-1]}$ for all $\ell\ge1$.

\begin{proposition}\label{prop:limval}
For all $w\in \adh(L)$, we have
\[\lim_{n\to+\infty}\frac{\val_S(w[0,n-1])}{v_{q_0}(n)}=\alpha_w.\]
\end{proposition}

\begin{proof}
Let $w\in \adh(L)$. 
For all $\ell$ and $n$ such that $n\ge\ell\ge1$, we have
\begin{equation}\label{eq:ineq}
\alpha_{w[0,\ell-1],n}
\le\frac{\val_S(w[0,n-1])}{v_{q_0}(n)}
<\alpha_{w[0,\ell-1],n}
+\frac{u_{q_0\cdot w[0,\ell-1]}(n-\ell)}{v_{q_0}(n)}.
\end{equation}
Let $\varepsilon>0$.
For all $\ell\ge1$, there exists $N(\ell)\ge\ell$ 
such that for all $n\ge N(\ell)$, we have
\[\alpha_{w[0,\ell-1]}-\frac{\varepsilon}{2}
<\frac{\val_S(w[0,n-1])}{v_{q_0}(n)}
<\alpha_{w[0,\ell-1]}+r_{w[0,\ell-1]}+\frac{\varepsilon}{2}.\]
By Hypothesis (H3) and Lemma~\ref{lem:a_w}, 
there exists also $k\in\mathbb{N}$ such that for all $\ell\ge k$, 
\[r_{w[0,\ell-1]}<\frac{\varepsilon}{2} 
\hspace{0.5cm} \text{ and } \hspace{0.5cm}
0<\alpha_w-\alpha_{w[0,\ell-1]}<\frac{\varepsilon}{2}.\]
It follows that for all $n\ge N(k)$,
\[\alpha_w-\varepsilon
<\alpha_{w[0,k-1]}-\frac{\varepsilon}{2}
<\frac{\val_S(w[0,n-1])}{v_{q_0}(n)}
<\alpha_w+\varepsilon\]
and the conclusion follows.
\end{proof}

The preceding proposition allows us to define the $S$-value of an infinite word in~$\adh(L)$. 

\begin{definition}
The application $\val_S\colon \adh(L)\to[s_0,1] \colon w\mapsto \alpha_w$ is called the {\em $S$-value} function.
\end{definition}

\begin{proposition}\label{lem:<}
If $w,z\in \adh(L)$ are such that $w$ is lexicographically less than $z$, then $\val_S(w)\le\val_S(z)$.
\end{proposition}

\begin{proof}
Let $w,z\in \adh(L)$. We deduce from \eqref{I_ya} that
if $k:=\inf\{i\in\mathbb{N}\,|\,w[i]<z[i]\}$,
then $\forall \ell\ge k$, we have $\alpha_{w[0,\ell-1]}\le \alpha_{z[0,\ell-1]}$ 
and the proposition holds.
\end{proof}

Recall now a result from \cite{BB}.

\begin{lemma}
If $K$ is an infinite language over a totally ordered alphabet,
then $\adh(K)$ contains a minimal element for the lexicographical ordering.
\end{lemma}

This leads to the following definition.

\begin{definition}
For all $y\in \centre(L)$, $m_y$ (resp. $M_y$) 
denotes the least (resp. greater) word in $\adh(L)$ in the lexicographical ordering 
having $y$ as a prefix.
\end{definition}

Notice that for all $y\in \centre(L)$,  we have $m_y=wv$ (resp. $M_y=wu$), 
where $u$ (resp. $v$) is the minimal (resp. maximal) 
word in $\adh(y^{-1}L)$ for the lexicographical ordering.

\begin{example}
Continuing Example \ref{ex:num}, 
we have $m_{aab}=aaba^\omega$ and $M_{aab}=aabb(ab)^\omega$.
\end{example}

\begin{lemma}\label{lem:mM}
For all $y\in \centre(L)$, one has
\[\val_S(m_y)=\alpha_y \;\text{ and }\;\val_S(M_y)=\alpha_y+r_y.\]
\end{lemma}

\begin{proof}
Let $y\in \centre(L)$. From \eqref{I_ya}, we get that for all $\ell\ge|y|$, 
$\alpha_{m_y[0,\ell-1]}=\alpha_y$ 
and $\alpha_{M_y[0,\ell-1]}+r_{M_y[0,\ell-1]}=\alpha_y+r_y$.
Therefore, we obtain that for all $\ell\ge|y|$,
\begin{align*}
\alpha_y \le &\val_S(m_y)\le\alpha_y+r_{m_y[0,\ell-1]},\\ 
\alpha_y+r_y-r_{M_y[0,\ell-1]} \le &\val_S(M_y) \le\alpha_y+r_y.
\end{align*}
We conclude by using Hypothesis (H3).
\end{proof}

\begin{proposition}
The $S$-value function is uniformly continuous. 
\end{proposition}

\begin{proof}
Let $w,z\in \adh(L)$. 
Assume that $d(w,z)=2^{-\ell}$. 
Then $w[0,\ell-1]=z[0,\ell-1]$ and, in view of Lemma~\ref{lem:mM},
the $S$-values $\val_S(w)$ and $\val_S(z)$ belong to $I_{w[0,\ell-1]}$. 
Thus $|\val_S(w)-\val_S(z)|\le r_{w[0,\ell-1]}\to 0$ 
as $\ell\to+\infty$ by Hypothesis (H3).
The conclusion follows.
\end{proof}

Using Lemma~\ref{lem:mM}, we are able to give an expresssion of 
the $S$-value of a word in $\adh(L)$. 

\begin{proposition}
For all $w\in \adh(L)$, 
\[\val_S(w)=s_0+\sum_{i=0}^{+\infty}\sum_{a<w[i]}r_{w[0,i-1]a}.\]
\end{proposition}

\begin{proof}
Let $w\in \adh(L)$. Using \eqref{I_ya}, we get that for all $n\ge1$,  
\begin{eqnarray*}
\alpha_{w[0,n-1]}
&=&s_0+\sum_{\substack{x<w[0,n-1]\\x\in \centre(L)\cap\Sigma^n}}r_x\\
&=&s_0+\sum_{i=0}^{n-1}\sum_{a<w[i]}\sum_{|y|=n-i-1}r_{w[0,i-1]ay}\\
&=&s_0+\sum_{i=0}^{n-1}\sum_{a<w[i]}r_{w[0,i-1]a}.
\end{eqnarray*}
Letting $n$ tend to infinity in the latter equality, 
we get the expected result.
\end{proof}

The following proposition links together the framework of \cite{LR2}, 
where are mainly considered converging sequences of words,
and the framework that has been developed in the present section 
to represent real numbers. 

\begin{proposition}
Let $K$ be a language over a totally ordered alphabet $(\Sigma,<)$ 
such that its prefix-closure $\pref(K)$ satisfies Hypotheses (H1), (H2), and (H3), 
and let $S=(\pref(K),\Sigma,<)$ be the abstract numeration system built on $\pref(K)$.
If $(w^{(n)})_{n\ge0}\in K^{\mathbb{N}}$ 
is a sequence of words such that $w^{(n)}\to w$, then we have
\[\lim_{n\to+\infty}\frac{\val_S(w^{(n)})}{v_{q_0}(|w^{(n)}|)}=\alpha_w.\]
\end{proposition}

\begin{proof}
Let $(w^{(n)})_{n\ge0}\in K^{\mathbb{N}}$ be 
a sequence of words such that $w^{(n)}\to w$.
Thanks to Lemma~\ref{lem:adh}, this implies that $\pref(w)\subseteq\pref(K)$.
For any $\ell\ge1$, 
there exists $N(\ell)\ge\ell$ such that for all $n\ge N(\ell)$,
$w^{(n)}[0,\ell-1]=w[0,\ell-1]$.
Then in view of \eqref{eq:valpref} and \eqref{eq:ineq}, 
for all $\ell\ge1$ and for all $n\ge N(\ell)$, we have 
\[\left|\frac{\val_S\left(w[0,|w^n|-1]\right)}{v_{q_0}(|w^n|)}-
\frac{\val_S\left(w^{(n)}\right)}{v_{q_0}(|w^n|)}\right|
\le \frac{u_{q_0\cdot w[0,\ell-1]}(|w^n|-\ell)}{v_{q_0}(|w^n|)}.\]
Let $\varepsilon>0$. 
By Hypothesis (H2), for all $\ell\ge1$, 
there exists $M(\ell)\ge\ell$ such that for all $n\ge M(\ell)$,
\[\frac{u_{q_0\cdot w[0,\ell-1]}(|w^n|-\ell)}{v_{q_0}(|w^n|)}
<r_{w[0,\ell-1]}+\frac{\varepsilon}{2}.\]
By Hypothesis (H3), there exists $k\in\mathbb{N}$ 
such that for all $\ell\ge k$, $r_{w[0,\ell-1]}<\frac{\varepsilon}{2}$.
Then for all $n\ge \max(N(k),M(k))$, we have
\[\left|\frac{\val_S\left(w[0,|w^n|-1]\right)}{v_{q_0}(|w^n|)}-
\frac{\val_S\left(w^{(n)}\right)}{v_{q_0}(|w^n|)}\right|<\varepsilon.\]
\end{proof}

To conclude this section, we recall some results from \cite{BB} interesting for our study.

\begin{proposition}
If $K$ is an infinite algebraic language over a totally ordered alphabet, 
then the minimal word of $\adh(K)$ is ultimately periodic
and can be effectively computed.
\end{proposition}

\begin{definition}
Let $K$ be a language over a totally ordered alphabet. 
The {\em minimal language of $K$}, denoted by $\min(K)$ 
is the language of the smallest words of each length for the lexicographical ordering:
\[\min(K)=\{w\in K\,|\,\forall z\in K, |w|=|z|\Rightarrow w<_{\text{lex}}z\}.\]
\end{definition}

\begin{proposition}
If $K$ is an infinite language such that $K=\centre(K)$, 
then we have $\min(K)=\pref(m_\varepsilon)$.
\end{proposition}

\begin{corollary}
If $K$ is an infinite algebraic language such that $K=\centre(K)$, 
then $\pref(m_\varepsilon)$ is a regular language.
\end{corollary}

Of course, all these results can be adapted to the case 
of the maximal word of the adherence of a language.

Transposed to the context of this paper, 
these results can be related to synctatical properties 
of the endpoints of the intervals $I_y$, for $y\in\centre(L)$.

\begin{corollary}\label{cor:endpoints}
Assume that the language $L$ is algebraic. 
Then for all $y \in\centre(L)$, 
the infinite words $m_y$ and $M_y$ are ultimately periodic.
\end{corollary}

Notice that in general, there exist ultimately periodic representations 
that are not endpoints of any interval $I_y$, where $y\in\centre(L)$.
For instance, in the integer base $10$ numeration system, 
we have that the representation of $\frac{1}{3}$ is $0.33333\cdots$ 
and $\frac{1}{3}$ is not the endpoint of any interval of the form 
$\left[\frac{k}{10^\ell},\frac{k+1}{10^\ell}\right]$, 
where $\ell\ge1$ and $k\in[\![0,10^\ell-1]\!]$.

\section{Applications}\label{sec:appl}

In this section, we apply our techniques to three examples to represent
real numbers in situations that were not settled in \cite{LR2}.
The first one shows how it can be easier to consider 
the prefix-closure of the language instead of the language itself.

\begin{example}
Consider again the language  
$L=\{w\in\{a,b\}^*\,|\, \left| |w|_a-|w|_b \right|\le 1\}$ 
of Example~\ref{ex:GANS}. 
This language is not prefix-closed. 
We have $\pref(L)=\{a,b\}^*$, which is of course a regular language.  
For the abstract numeration system $S=(\pref(L),\{a,b\},a<b)$,  
the hypotheses (H1), (H2) and (H3) are trivially satisfied. 
More precisely, for all $w\in\{a,b\}^*$, we have $r_w=2^{-|w|-1}$. 
Using the same notation as in Example~\ref{ex:GANS}, we have 
\[\lim_{n\to+\infty}\frac{v_0(n-1)}{v_0(n)}=\frac{1}{2}.\]
Therefore, we represent the interval $[\frac{1}{2},1]$.
For all $\ell\ge1$, $\centre(L)\cap\Sigma^\ell=\{a,b\}^\ell$ and 
the intervals corresponding to words of length $\ell$ 
are exactly the intervals $\left[\frac{k}{2^\ell},\frac{k+1}{2^\ell}\right]$, 
for any $k\in[\![0,2^\ell-1]\!]$.
\end{example}

The second example illustrates the case of 
a non-regular language with a non-regular prefix-language. 

\begin{example}\label{ex:dyck}
The {\em Dyck language} is the language 
\[D:=\{w\in\{a,b\}^*|\,|w|_a=|w|_b \text{ and }\forall u\in \pref(w),\,|u|_b\ge|u|_a\}\]
of the well-parenthesized words over two letters.
Its (infinite) minimal automaton $\mathcal{A}_D=\{Q,q_0,\{a,b\},\delta,\{q_0\})$ 
is represented in Figure~\ref{aut:dyck}. 
For each $m\ge0$, define $d_m={(a^m)}^{-1}D=\{w\in\{a,b\}^*|\,a^mw\in D\}$ and $d_{-1}=\emptyset$, 
so that $Q=\{d_m\,|\,m\ge0\}\cup\{d_{-1}\}$. 
Notice that in Figure \ref{aut:dyck}, the states $d_m$ are simply denoted by $m$.

\begin{figure}[htbp]
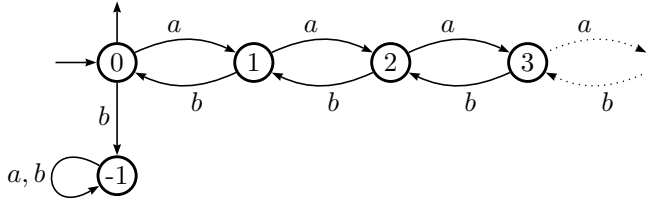

        \centering
\VCDraw{%
        \begin{VCPicture}{(0,-0.2)(12,3.8)}
 \State[0]{(0,2.5)}{I}
 \State[1]{(3,2.5)}{Q}
 \State[2]{(6,2.5)}{R}
 \State[3]{(9,2.5)}{S}
 \State[$-$1]{(0,0)}{P}
\HideState
\State[4]{(12,2.5)}{T}
\Initial[w]{I}
\Final[n]{I}
\VArcL{arcangle=25}{I}{Q}{a}
\EdgeR{I}{P}{b}
\VArcL{arcangle=25}{Q}{I}{b}
\VArcL{arcangle=25}{Q}{R}{a}
\VArcL{arcangle=25}{R}{Q}{b}
\VArcL{arcangle=25}{R}{S}{a}
\VArcL{arcangle=25}{S}{R}{b}
\LoopW[.5]{P}{a,b}
\ChgEdgeLineStyle{dotted}
\VArcL{arcangle=25}{S}{T}{a}
\VArcL{arcangle=25}{T}{S}{b}
\end{VCPicture}
}
        \caption{The minimal automaton of $D$.}
        \label{aut:dyck}
\end{figure}

\noindent
It has been proved in \cite{Marion} that for all $m\ge0$,
\[u_{d_m}(n)=
\left\{
\begin{array}{ll}
0 & \text{ if } n<m \text { or } m\not\equiv n\mod 2,\\
\frac{m+1}{n+1}\binom{n+1}{\frac{n-m}{2}} &\text{ if }n\ge m \text{ and }m\equiv n\mod 2.
\end{array}
\right.
\]
By Stirling's formula, we get that for all $m\ge0$,
\begin{gather}
\label{eq:u_2m_2n}
        u_{d_{2m}}(2n)\sim\frac{2m+1}{\sqrt{\pi}}\,n^{-\frac{3}{2}} 4^n \;(n\to +\infty),\\     
\label{eq:u_2m+1_2n+1}
        u_{d_{2m+1}}(2n+1)\sim \frac{2(2m+2)}{\sqrt{\pi}}\,n^{-\frac{3}{2}} 4^n \;(n\to +\infty).
\end{gather}
                
The Dyck language is not prefix-closed. 
Hence we consider the abstract numeration system 
$S=(P,\{a,b\},a<b)$ built on the language 
\[P:=\pref(D)=\{w\in\{a,b\}^*|\,\forall u\in \pref(w),\,|u|_b\ge|u|_a\}\]
of the prefixes of the Dyck words. 
The (infinite) minimal automaton of $P$ is 
$\mathcal{A}_{P}=(Q,q_0,\{a,b\},\delta,F)$. 
It is represented in Figure~\ref{aut:prefdyck}. 
Since the minimal automaton $\mathcal{A}_{P}$ of $P$ 
and the minimal automaton $\mathcal{A}_{D}$ of $D$ are nearly the same, 
we rename the states of $\mathcal{A}_{P}$ by $p_m:=d_m$. 
Hence the $u_{d_m}$'s denotes the complexity functions of $\mathcal{A}_{D}$ 
and the $u_{p_m}$'s denotes the complexity functions of $\mathcal{A}_{P}$.
By Proposition~\ref{prop:linfty}, $\adh(P)=\adh(D)$ is uncountable
and Hypothesis (H1) is satisfied.

\begin{figure}[htbp]
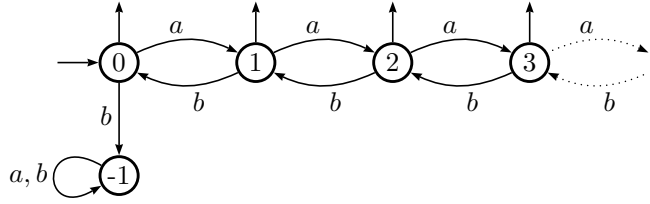

        \centering
\VCDraw{%
        \begin{VCPicture}{(0,-0.2)(12,3.8)}
 \State[0]{(0,2.5)}{I}
 \State[1]{(3,2.5)}{Q}
 \State[2]{(6,2.5)}{R}
 \State[3]{(9,2.5)}{S}
 \State[$-$1]{(0,0)}{P}
\HideState
\State[4]{(12,2.5)}{T}
\Initial[w]{I}
\Final[n]{I}
\Final[n]{Q}
\Final[n]{R}
\Final[n]{S}
\VArcL{arcangle=25}{I}{Q}{a}
\EdgeR{I}{P}{b}
\VArcL{arcangle=25}{Q}{I}{b}
\VArcL{arcangle=25}{Q}{R}{a}
\VArcL{arcangle=25}{R}{Q}{b}
\VArcL{arcangle=25}{R}{S}{a}
\VArcL{arcangle=25}{S}{R}{b}
\LoopW[.5]{P}{a,b}
\ChgEdgeLineStyle{dotted}
\VArcL{arcangle=25}{S}{T}{a}
\VArcL{arcangle=25}{T}{S}{b}
\end{VCPicture}
}
        \caption{The minimal automaton of $\pref(D)$.}
        \label{aut:prefdyck}
\end{figure}    

\noindent
Observe that for all $m\ge0$,
\[u_{p_m}(n)=
 \left\{
\begin{array}{ll}
2^n & \text{ if } n\le m,\\
2u_{p_m}(n-1)-u_{d_m}(n-1) & \text{ if }n> m.\\
\end{array}
\right.
\] 
Hence we get that for all $m\ge0$,
\[u_{p_m}(n)=
 \left\{
\begin{array}{ll}
2^n & \text{ if } n\le m,\\
2^n-\sum_{i=m}^{n-1}u_{d_{m}}(i)\,2^{n-i-1} & \text{ if }n>m.
\end{array}
\right.
\]
We have that for all $m\ge0$,
\begin{gather}\label{eq:sim} 
u_{p_m}(2n)\sim\frac{m+1}{\sqrt{\pi}}n^{-\frac{1}{2}} 4^n\; (n\to +\infty),\\
\label{eq:sim2}
u_{p_m}(2n+1)\sim v_{p_m}(2n)\sim \frac{2(m+1)}{\sqrt{\pi}}n^{-\frac{1}{2}} 4^n\; (n\to +\infty),\\
\label{eq:sim3}
v_{p_m}(2n+1)\sim\frac{4(m+1)}{\sqrt{\pi}}n^{-\frac{1}{2}} 4^n\; (n\to +\infty).
\end{gather}
We prove only \eqref{eq:sim} since the same techniques 
can be applied to obtain \eqref{eq:sim2} and \eqref{eq:sim3}.
Let us first show that for all $m\ge0$, 
we have 
\begin{equation}\label{eq:sum}
\sum_{i=m}^{+\infty}u_{d_{2m}}(2i)\,4^{-i}=2 
\hspace{0.5cm}\text{ and }\hspace{0.5cm}
\sum_{i=m}^{+\infty}u_{d_{2m+1}}(2i+1)\,4^{-i}=4.
\end{equation}
We compute only the first sum, the second one can be treated in similar way.
In view of \eqref{eq:u_2m_2n} and \cite[Ch. V.4, Prop. 2]{bourbaki}, for all $m\ge0$,
we have
\[\sum_{i=n}^{+\infty}u_{d_{2m}}(2i)4^{-i}\sim 
\frac{2m+1}{\sqrt{\pi}}\sum_{i=n}^{+\infty}i^{-\frac{3}{2}}\; (n\to +\infty) \]
and the series 
\[\sum_{i=m}^{+\infty}u_{d_{2m}}(2i)4^{-i}\]
is convergent.
Consequently, for all $m\ge0$, the series
\[\sum_{i=m}^{+\infty}u_{d_{2m}}(2i)\,z^i \]
is uniformly convergent over $\{z\in\mathbb{C}\,|\,|z|\le\frac{1}{4}\}$ 
because for all $q\ge p\ge m$, we have
\[\sup_{|z|\le\frac{1}{4}}\left|\sum_{i=p}^{q}u_{d_{2m}}(2i)\,z^i\right|
\le\sum_{i=p}^{q}u_{d_{2m}}(2i)4^{-i}.\] 
Then observe that for all $m\ge0$ and $i\ge m$ 
such that $i\equiv m\mod 2$, we have 
\begin{eqnarray*}
u_{d_{m}}(i) &=& \text{Card}\{w^{(0)}bw^{(1)}b\cdots bw^{(m)}\mid
                \forall j\in[\![0,m]\!],\,w^{(j)}\in D,\, \sum_{j=0}^m |w^{(j)}|=i-m\}\\
             &=& \sum_{\ell_0+\cdots+\ell_m=\frac{i-m}{2}}
                \left(\prod_{j=0}^m \mathcal{C}_{\ell_j}\right)
                =\left[z^{\frac{i-m}{2}}\right]
                \left(\sum_{n=0}^{+\infty}\mathcal{C}_n\,z^n\right)^{m+1}
\end{eqnarray*}
where $\mathcal{C}_n:=u_{d_0}(2n)=\frac{1}{2n+1}\binom{2n+1}{n}$
is the $n$th Catalan number \cite{GKP} 
and $[z^n]f$ is the coefficient of $z^n$ in the power series $f$. 
It is well known that
\[\sum_{n=0}^{+\infty}\mathcal{C}_n\,z^n=\frac{1-\sqrt{1-4z}}{2z}\]
for $|z|<\frac{1}{4}$. Hence we get that for all $m\ge0$,
\[\sum_{i=m}^{+\infty}u_{d_{2m}}(2i)\,z^i
=z^m\left(\sum_{n=0}^{+\infty}\mathcal{C}_n\,z^n\right)^{2m+1}
=\frac{{(1-\sqrt{1-4z})}^{2m+1}}{2\cdot\,4^m z^{m+1}}\]
Therefore, we obtain the desired first sum of \eqref{eq:sum}
by letting $z$ tend to $\frac{1}{4}$ in the corresponding formula. 
We now come back on \eqref{eq:sim}.
For all $0\le m<n$, we have
\[u_{p_{2m}}(2n)
= 4^n-\frac{1}{2}\sum_{i=m}^{n-1}u_{d_{2m}}(2i)\,4^{n-i}
= \frac{1}{2}\,4^n \sum_{i=n}^{+\infty}u_{d_{2m}}(2i)\,4^{-i}\]
and 
\[u_{p_{2m+1}}(2n) 
= 4^n-\frac{1}{4}\sum_{i=m}^{n-1}u_{d_{2m+1}}(2i+1)\,4^{n-i}
= \frac{1}{4}\,4^n \sum_{i=n}^{+\infty}u_{d_{2m+1}}(2i+1)\,4^{-i}.\]
Notice that 
$\sum_{i=n}^{+\infty} i^{-\frac{3}{2}} \sim 2 n^{-\frac{1}{2}}$.
Finally we obtain that for all $m\ge0$,
\[u_{p_{2m}}(2n)\sim \frac{2m+1}{\sqrt{\pi}}n^{-\frac{1}{2}}4^n
\hspace{0.5cm}\text{ and }\hspace{0.5cm}
u_{p_{2m+1}}(2n)\sim \frac{2m+2}{\sqrt{\pi}}n^{-\frac{1}{2}}4^n,\]
proving \eqref{eq:sim}. 

Let us now verify that the language $P$ satisfies our three hypotheses. 
From the previous reasoning, we get that for all $m\ge0$ and all $\ell\ge 0$,
\[\lim_{n\to+\infty} \frac{u_{p_m}(n-\ell)}{v_{p_0}(n)}=(m+1)\,2^{-\ell-1}.\] 
For all $w\in P$, $r_w:=(m_w+1)\,2^{-|w|-1}$ 
where $m_w$ is defined by $p_0\cdot w=p_{m_w}$ 
and for all $w\not\in P$, $r_w:=0$. 
Hence Hypothesis (H2) is satisfied. 
Let now $w\in \adh(D)$. 
Observe that $m_{w[0,\ell-1]}\le\ell$ for all $\ell\ge1$. 
Therefore, for all $w\in \adh(D)$, 
we have $r_{w[0,\ell-1]}\le (\ell+1)2^{-\ell-1}\to0$ as $\ell\to\infty$ 
and Hypothesis (H3) is satisfied. 

Since 
\[\lim_{n\to+\infty} \frac{v_{p_0}(n-1)}{v_{p_0}(n)}=\frac{1}{2},\]
we represent the interval $[\frac{1}{2},1]$.
We have $\centre(D)\cap\Sigma^\ell=P\cap\{a,b\}^\ell$. 
Any word of $P$ begins with $a$, so that $I_a=[\frac{1}{2},1]$.
We have $\centre(D)\cap\Sigma^2=\{aa,ab\}$ and $I_a$ is partitioned into two subintervals: 
\[I_{aa}=\left[\frac{1}{2},\frac{7}{8}\right] 
\hspace{0.5cm} \text{ and } \hspace{0.5cm}
I_{ab}=\left[\frac{7}{8},1\right]\] 
Then $\centre(D)\cap\Sigma^3=\{aaa,aab,aba\}$. 
Thus $I_{ab}=I_{aba}$ and $I_{aa}$ is partitioned into two new subintervals 
\[I_{aaa}=\left[\frac{1}{2},\frac{3}{4}\right],\;
I_{aab}=\left[\frac{3}{4},\frac{7}{8}\right],\;I_{aba}=\left[\frac{7}{8},1\right].\]
Then $\centre(D)\cap\Sigma^4=\{aaaa,aaab,aaba,aabb,,abaa,abab\}$ and we get 
\begin{gather*}
I_{aaaa}=\left[\frac{1}{2},\frac{21}{32}\right],\;
I_{aaab}=\left[\frac{21}{32},\frac{3}{4}\right],\; 
I_{aaba}=\left[\frac{3}{4},\frac{27}{32}\right],\\
I_{aabb}=\left[\frac{27}{32},\frac{7}{8}\right],\;
I_{abaa}=\left[\frac{7}{8},\frac{31}{32}\right],\;
I_{abab}=\left[\frac{31}{32},1\right].
\end{gather*}

As stated by Corollary~\ref{cor:endpoints}, since the language $D$ is algebraic, 
for all $y\in\centre(D)$, the representations of the endpoints 
of the interval $I_y$ are ultimately periodic.
Let $Q_x$ denotes the set of all the representations of $x$. 
We have $Q_{\frac{1}{2}}=\{a^\omega\}$ and $Q_1=\{(ab)^\omega\}$.
Now let $x\in(\frac{1}{2},1)$ be an endpoint of some interval, 
i.e., $x=\inf I_w=\sup I_z$ for some
$w,z\in \centre(D)\cap\Sigma^\ell$ with $\ell\ge0$. We have
$Q_x=\{\bar{w}(ab)^\omega,za^\omega\}$,
where $\bar{w}$ is the smallest Dyck word having $w$ as a prefix.   
\end{example}

The third example illustrates the case 
of a generalized abstract numeration systems generating 
endpoints of the intervals $I_y$ having no ultimately 
periodic $S$-representations. 
It also shows that our methods for representing reals 
generalize the ones involved to represent reals in the $\frac{3}{2}$-number 
system and by extension the rational base number systems as well.

\begin{example}\label{ex:3/2}
Consider the language $L:=L_{\frac{3}{2}}$ recognized by the deterministic automaton 
$\mathcal{A}=(\mathbb{N}\cup\{-1\},0,\{0,1,2\},\delta,\mathbb{N})$ 
where the transition function $\delta$ 
is defined as follows: 
$\delta(n,a)=\frac{1}{2}(3n+a)$ if $n\in\mathbb{N}$ and $a\in\{0,1,2\}$ are such that $\frac{1}{2}(3n+a)\in\mathbb{N}$ and $\delta(n,a)=-1$ otherwise.  
This language has been introduced and studied in \cite{AFS}. 
In particular, it has been shown that the automaton 
$\mathcal{A}$ is the minimal automaton of $L$, that $L$ is a 
non-algebraic prefix-closed language and that $\adh(L)$ is uncountable. 
Moreover, no element of $\adh(L)$ is ultimately periodic.
The corresponding trim minimal automaton is depicted in Figure~\ref{aut:3/2}, 
where all states are final. 
\begin{figure}[htbp]
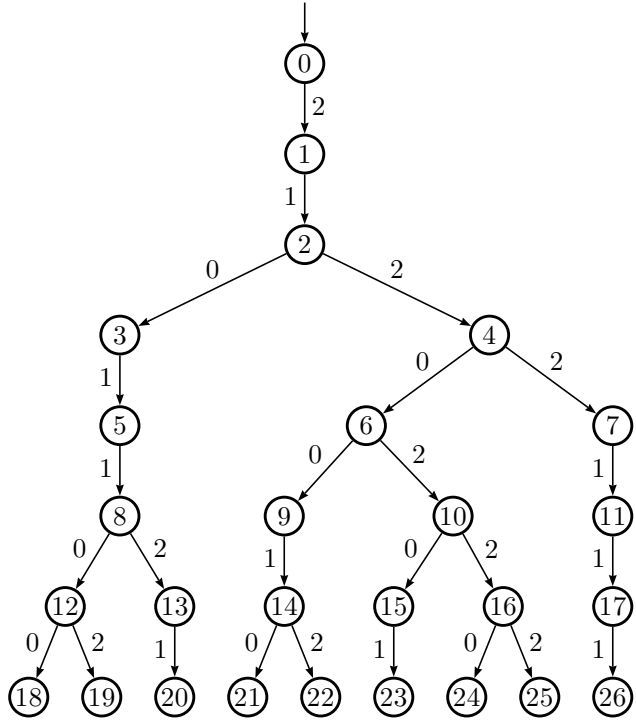

        \centering
\VCDraw{%
        \begin{VCPicture}{(0,0)(12.8,15)}
 \State[0]{(6.05,14)}{0}
 \State[1]{(6.05,12)}{1}
 \State[2]{(6.05,10)}{2}
\State[3]{(2,8)}{3}
 \State[4]{(10.1,8)}{4}
\State[5]{(2,6)}{5}
 \State[6]{(7.4,6)}{6}
 \State[7]{(12.8,6)}{7}
\State[8]{(2,4)}{8}
 \State[9]{(5.6,4)}{9}
 \State[10]{(9.3,4)}{10}
 \State[11]{(12.8,4)}{11}
\State[12]{(0.8,2)}{12}
 \State[13]{(3.2,2)}{13}
 \State[14]{(5.6,2)}{14}
 \State[15]{(8,2)}{15}
 \State[16]{(10.4,2)}{16}
 \State[17]{(12.8,2)}{17}
\State[18]{(0,0)}{18}
 \State[19]{(1.6,0)}{19}
 \State[20]{(3.2,0)}{20}
 \State[21]{(4.8,0)}{21}
 \State[22]{(6.4,0)}{22}
 \State[23]{(8,0)}{23}
 \State[24]{(9.6,0)}{24}
 \State[25]{(11.2,0)}{25}
 \State[26]{(12.8,0)}{26}
\Initial[n]{0}
\EdgeL{0}{1}{2}
\EdgeR{1}{2}{1}
\EdgeR{2}{3}{0}
\EdgeL{2}{4}{2}
\EdgeR{3}{5}{1}
\EdgeR{4}{6}{0}
\EdgeL{4}{7}{2}
\EdgeR{5}{8}{1}
\EdgeR{6}{9}{0}
\EdgeL{6}{10}{2}
\EdgeR{7}{11}{1}
\EdgeR{8}{12}{0}
\EdgeL{8}{13}{2}
\EdgeR{9}{14}{1}
\EdgeR{10}{15}{0}
\EdgeL{10}{16}{2}
\EdgeR{11}{17}{1}
\EdgeR{12}{18}{0}
\EdgeL{12}{19}{2}
\EdgeR{13}{20}{1}
\EdgeR{14}{21}{0}
\EdgeL{14}{22}{2}
\EdgeR{15}{23}{1}  
\EdgeR{16}{24}{0}
\EdgeL{16}{25}{2}
\EdgeR{17}{26}{1}
\end{VCPicture}
}
        \caption{First levels of the trim minimal automaton of $L_{\frac{3}{2}}$.}
        \label{aut:3/2}
\end{figure}

Let $(G_n)_{n\ge0}$ be the sequence of integers defined by:
\[G_0=1 \; \text{ and } \; \forall n\in\mathbb{N},\;
G_{n+1}:=\left\lceil\frac{3}{2}G_n\right\rceil.\]
From \cite{AFS}, we find 
\[u_0(0)=1 \; \text{ and } \; 
\forall n\in\mathbb{N},\;u_0(n+1)=G_{n+1}-G_{n}.\]
It has been shown in \cite{AFS} that for all $n\ge0$, 
$G_n=\lfloor K\left(\frac{3}{2}\right)^n\rfloor$, 
where $K:=K(3)=1.6222705\cdots$ is the constant discussed 
in \cite{MR1108748,MR1617400,Stephan}.  
Consider now the abstract numeration system 
$S=(L,\{0,1,2\},0<1<2)$ built on this language. 
From \cite{AFS}, we know that for all $w\in L$, 
\[\val_S(w)
=\frac{1}{2}\sum_{i=0}^{|w|-1} w[i]\left(\frac{2}{3}\right)^{|w|-1-i}.\]
Consequently, for all $w\in \adh(L)$, we have 
\[\val_S(w)
=\frac{1}{3K}\sum_{i=0}^{+\infty} w[i]\left(\frac{2}{3}\right)^i.\]
Now let us verify that $L$ satisfies Hypothesis (H2) and (H3). 
Recall that, for all $x\in L$, $M_x$ (resp. $m_x$) denotes the maximal 
(resp. minimal) word in $\adh(L)$
for the lexicographic ordering having $x$ as a prefix. 
We have that, for all $x\in L$,
\begin{eqnarray*}
r_x=|I_x| &=& \val_S(M_x)-\val_S(m_x)\\
        &=& \frac{1}{3K}\sum_{i=|x|}^{+\infty}
                (M_x[i]-m_x[i])\left(\frac{2}{3}\right)^i\\
        &=& \frac{1}{3K}\left(\frac{2}{3}\right)^{|x|}\,
                \sum_{i=0}^{+\infty}(M_x [i+|x|]-m_x [i+|x|])\left(\frac{2}{3}\right)^i\ge0
\end{eqnarray*}
and Hypothesis (H2) is satisfied. 
For all $x\in L$, since $M_x [i]-m_x[i]\le2$ for all $i\ge0$, 
we obtain from that
\[r_x \le \frac{2}{K}\left(\frac{2}{3}\right)^{|x|}\to 0 \text{ as } |x|\to+\infty.\]
Therefore, if $w\in \adh(L)$, then $\lim_{\ell\to+\infty}w[0,\ell-1]=0$ 
and Hypothesis (H3) is also satisfied.
\end{example}

\section*{Open poblems}
\begin{itemize}
 \item Find a necessary condition on any automaton 
recognizing a language $L$ so that the corresponding $\omega$-language $\adh(L)$ is uncountable.
 \item Let $D_2$ be the Dyck language for two kinds of parentheses. It is well-known that for every algebraic
language $L$, there exists a faithful sequential mapping $f$ such that $f(\adh(D_2))=\adh(f(D_2))=\adh(L)$, 
see \cite[Theorem 6]{BN} for details. Let $S$ and $T$ be abstract numeration systems built respectively on $\pref(D_2)$ and $\pref(L)$. Give a mapping $g$ such that the following diagram commutes.
\[\xymatrix{ \adh(D_2) \ar[r]^f \ar[d]_{\val_S} & \adh(L) \ar[d]^{\val_T}\\
                [s_0,1] \ar[r]_{g} & [t_0,1]}\]
\end{itemize}

\section*{Acknowledgements}
We thank Professor J.-P. Schneiders for fruitful discussions. 
This work was initiated during the post-doctoral stay of the second author at the University of Liege, 
thanks to a federal research allowance.

\bibliography{bibliographie}

\begin{thebibliography}{SYZS92}

\bibitem[AFS08]{AFS}
S.~Akiyama, C.~Frougny, and J.~Sakarovitch.
\newblock Powers of {R}ationals {M}odulo 1 and {R}ational {B}ase {N}umber
  {S}ystems.
\newblock {\em Israel J. Math.}, 168:53--91, 2008.

\bibitem[BB97]{BB}
J.~Berstel and L.~Boasson.
\newblock The set of minimal words of a context-free language is context-free.
\newblock {\em J. Comput. System Sci.}, 55(3):477--488, 1997.

\bibitem[BN80]{BN}
L.~Boasson and M.~Nivat.
\newblock Adherences of languages.
\newblock {\em J. Comput. System Sci.}, 20(3):285--309, 1980.

\bibitem[Bou07]{bourbaki}
N.~Bourbaki.
\newblock {\em {F}onctions d'une variable r\'eelle.}
\newblock Springer Berlin Heidelberg, 2007.

\bibitem[DT89]{DT}
J.-M. Dumont and A.~Thomas.
\newblock Syst\`{e}mes de num\'eration et fonctions fractales relatifs aux
  substitutions.
\newblock {\em Theoret. Comput. Sci.}, 65(2):153--169, 1989.

\bibitem[Eil74]{Ei}
S.~Eilenberg.
\newblock {\em Automata, {L}anguages, and {M}achines}, volume~A.
\newblock Academic Press, New York, 1974.
\newblock Pure and Applied Mathematics, Vol. 58.

\bibitem[GKP94]{GKP}
Ronald~L. Graham, Donald~E. Knuth, and Oren Patashnik.
\newblock {\em Concrete {M}athematics}.
\newblock Addison-Wesley Publishing Company, Reading, MA, second edition, 1994.
\newblock A Foundation for Computer Science.

\bibitem[HH97]{MR1617400}
L.~Halbeisen and N.~Hungerb{\"u}hler.
\newblock The {J}osephus {P}roblem.
\newblock {\em J. Th\'eor. Nombres Bordeaux}, 9(2):303--318, 1997.

\bibitem[LG08]{Marion}
M.~Le~Gonidec.
\newblock On {C}omplexity of {I}nfinite {W}ords {A}ssociated with {G}eneralized
  {D}yck {L}anguages.
\newblock {\em Theoret. Comput. Sci.}, 407(1-3):117--133, 2008.

\bibitem[Lot02]{Lot}
M.~Lothaire.
\newblock {\em Algebraic {C}ombinatorics on {W}ords}, volume~90 of {\em
  Encyclopedia of Mathematics and its Applications}.
\newblock Cambridge University Press, Cambridge, 2002.

\bibitem[LR01]{LR}
P.~B.~A. Lecomte and M.~Rigo.
\newblock Numeration {S}ystems on a {R}egular {L}anguage.
\newblock {\em Theory Comput. Syst.}, 34(1):27--44, 2001.

\bibitem[LR02]{LR2}
P.~B.~A. Lecomte and M.~Rigo.
\newblock On the {R}epresentation of {R}eal {N}umbers {U}sing {R}egular
  {L}anguages.
\newblock {\em Theory Comput. Syst.}, 35(1):13--38, 2002.

\bibitem[Niv78]{N}
M.~Nivat.
\newblock Sur les ensembles de mots infinis engendr\'es par une grammaire
  alg\'ebrique.
\newblock {\em RAIRO Inform. Th\'eor.}, 12(3):259--278, v, 1978.

\bibitem[OW91]{MR1108748}
A.~Odlyzko and H.~Wilf.
\newblock Functional {I}teration and the {J}osephus {P}roblem.
\newblock {\em Glasgow Math. J.}, 33(2):235--240, 1991.

\bibitem[Sak03]{Saka}
J.~Sakarovitch.
\newblock {\em \'Elements de th\'eorie des automates}.
\newblock Vuibert, Paris, 2003.

\bibitem[Ste03]{Stephan}
R.~Stephan.
\newblock On a {S}equence {R}elated to the {J}osephus {P}roblem.
\newblock http://arxiv.org/abs/math/0305348v1, 2003.

\bibitem[SYZS92]{Sz}
A.~Szilard, S.~Yu, K.~Zhang, and J.~Shallit.
\newblock Characterizing {R}egular {L}anguages with {P}olynomial {D}ensities.
\newblock In {\em Mathematical {F}oundations of {C}omputer {S}cience 1992
  ({P}rague, 1992)}, volume 629 of {\em Lecture Notes in Comput. Sci.}, pages
  494--503. Springer, Berlin, 1992.

\end{thebibliography}
\bibliographystyle{alpha}

\end{document}